\documentclass[iop]{emulateapj}
\usepackage{amsmath, lettrine}
\usepackage{placeins}
\usepackage{booktabs}
\usepackage{color}

\shorttitle{Dimensionless scaling from laboratory to astrophysics.}
\shortauthors{Cross, Reville \& Gregori}
\begin{document}



\title{Scaling of Magneto-Quantum-Radiative Hydrodynamic Equations: From Laser-produced plasmas to Astrophysics}



\author{J. E. Cross\altaffilmark{1,*}, B. Reville\altaffilmark{2}, and G. Gregori\altaffilmark{1}}

\altaffiltext{1}{Clarendon Laboratory, University of Oxford, Parks Road, Oxford OX1 3PU, United Kingdom}
\altaffiltext{2}{Centre for Plasma Physics, Queen's University Belfast, University Road, BT7 1NN, United Kingdom}
\altaffiltext{*}{j.e.cross@physics.ox.ac.uk}



\begin{abstract}
We introduce here the equations of magneto-quantum-radiative hydrodynamics. By rewriting them in a dimensionless form, we obtain a set of parameters that describe scale-dependent ratios of all the characteristic hydrodynamic quantities. We discuss how these dimensionless parameters relate to the scaling between astrophysical observations and laboratory experiments.  
\end{abstract}



\section{Introduction}
The study of astrophysical phenomena using laser-produced plasma is a growing field of research \citep{Remington1999,nmagnetic, savin2012, Meinecke2014}.
Modern laser facilities can deliver large amounts of energy in very short times, exceeding what is possible from more conventional techniques such as gas guns or pulsed power machines. Pressure near the laser spot (where most of the laser energy is deposited) can reach values in excess of tens of Mbar, and is comparable to the energy density of bound electrons in atoms. Under these conditions quantum processes and radiation diffusion can also become important.  Return currents, as well as steep density and pressure gradients produce magnetic fields \citep{HainesPPCF1986} which can modify the overall transport of charged particles. These large deposited energies then drive powerful shock waves into the ambient medium \citep{Foster2002,Robey2002, Klein2003,Hansen2005}. The process bears similarities with many astrophysical phenomena where energy is impulsively released in the interstellar medium, such as supernova remnants \citep{Chevalier1992}, Herbig-Haro flows \citep{hartigan1987} and accretion shocks \citep{Miniati2000}. 

Laboratory experiments offer a viable complementary approach to both astrophysical observations (by providing, for example, the means of directly measuring quantities of interest not accessible by observation) and numerical calculations, thus overcoming limitations in resolution, numerical viscosity and potentially addressing non-linear aspects of the dynamical evolution, and/or validating simulation codes. 

This is meaningful only if the relevant physics in the laboratory is related to the astrophysical object. We refer to this as a {\it similarity} relation between the two systems. The most obvious situation is one where the laboratory experiment reaches the exact conditions found in the astrophysical object. 
This has been exploited, for example, to study the equation of state of planetary interiors \citep{jeanloz2007} and other compact objects
\citep{Kritcher2008, GarciaSaiz2008}. However it is not always possible to reach the exact conditions which we are interested in, as the spatial, temporal and energy scales may be outside the range of what is directly reproducible in an experiment. A similarity relation still exists if we can show that the laboratory and astrophysical systems evolve in a way that the governing equations are invariant under a scale transformation; this requires the corresponding spatial, density, pressure, time, and so on, values in one system to be mapped onto the other system by multiplicative constants. This similarity can be obtained via fluid equations \citep{Livermore1999}, or even at the kinetic level \citep{nfscaling,Ryutov2012} under some conditions. This paper concerns the magneto-hydrodynamics (MHD) similarity, and provides a general framework to include effects arising from finite resistivity, thermal conduction, radiation diffusion and quantum non-locality. 

Fluid similarity has previously been discussed, quite extensively, by Ryutov and Falize \citep{Livermore1999, Ryutov2000, Ryutov2001, Ryutov2012, Falize2011, Falize2011a}.  On the other hand, only selected aspects of the full governing equations have been investigated in the previous work, {\it i.e.} viscous hydrodynamics, radiative effects, or resistive MHD. The aim of the present work is thus to bring all the different elements of the equations together in a simple conceptual form.  We consider here the most general form of the fluid equations including magnetic, radiative and quantum effects, which are therefore applicable to a wide range of cases.  By rewriting these equations in a dimensionless form we derive a set of characteristic ratios containing the details of the microscopic properties of the fluid at a given scale. Values of these ratios tells us how important local properties are in determining the overall fluid motion. 

We introduce the full set of fluid equations in \S2, and specialise them for the case of an optically thick plasma in \S3. In \S3 we also discuss the Bohm potential and the inclusion of quantum dynamics in the fluid model. This can become important for exotic matter, such as inside neutron stars or white dwarfs.  Section \S4 describes the dimensionless analysis, and in section \S5 we derive the dimensionless fluid equations for the optically thick case. The optically thin equations are instead given in \S6.  Section \S7 discusses the different dimensionless numbers and their relevance to experiments. In \S8 we compare some laboratory experiments with their astrophysical counterparts and apply similarity in the context of our dimensionless analysis.  We draw our conclusions in \S9.

\section{General Equations}
\label{GenEq}

While the equations of fluid dynamics are the same everywhere in the Universe, there is no guarantee that a laboratory fluid would behave in the same manner as an astrophysical fluid. The two systems will exhibit the same dynamics only under some specific conditions. In order to explicitly extract such relations, we first write the full set of MHD equations in presence of heat conduction, radiation diffusion and quantum effects:
\citep{zeldovich2002, drake2006high,McClarren2010,quantum}. 
We assume the plasma is described by a single fluid, but with appropriate transport coefficients that are 
derived from kinetic theory \citep{chapman}.

\subsection{Continuity Equation}
The equation for the conservation of mass is given by:
\begin{subequations}
\label{FDequations}
\begin{equation}
\label{Continuity}
{\partial \rho \over \partial t} + \nabla \cdot \rho \boldsymbol{u} = 0 ,
\end{equation}
where $\rho$ is the mass density, $t$ the time and $\boldsymbol{u}$ the fluid velocity.

\subsection{Momentum Equation}
The equation for conservation of momentum reads as:
\begin{equation}
\label{momentum}
\rho \left({\partial \boldsymbol{u} \over \partial t} + \boldsymbol{u} \cdot  \nabla \boldsymbol{u} \right) = -\nabla p +\Phi_{Bohm}+ \nabla \cdot \boldsymbol {\sigma}_{\nu} + \boldsymbol{F}_{EM} + f_{rad}, 
\end{equation}
where $p$ is the fluid (ram) pressure, $\Phi_{Bohm}$ the quantum Bohm potential, $\boldsymbol{\sigma}_v$ the stress tensor,  $\boldsymbol{F}_{EM}$ electromagnetic volume forces, and $f_{rad}$ the volume force of radiation on matter. This equation shows that, in the most general case, the momentum associated to a fluid element can change not only by the inertial term and the pressure gradient, but also due to exchange effects (the Bohm potential contribution), viscous drag, and radiative forces. Each one of these non-ideal terms will be discussed in detail in the following sections. 
We also note that in equation (\ref{momentum}) the radiation force on matter, $f_{rad}$, in its most general form, includes effects from absorption and scattering \citep{Shu1}.

\subsection{Energy Equation}

The equation for conservation of energy is:
\begin{multline}
\label{energy}
{\partial \over \partial t} \left( \rho \epsilon + {\rho u^{2} \over 2} + E_R\right) + \nabla \cdot \left[\rho \boldsymbol{u} \left(\epsilon + {u^{2} \over 2} \right) + p \boldsymbol{u} \right] \\= -\nabla \cdot \boldsymbol{H} - \boldsymbol{J} \cdot \boldsymbol{E} + \Phi_{Bohm} \cdot \boldsymbol{u} - f_{rad} \cdot \boldsymbol{u}, 
\end{multline}
where  $\epsilon$ is the specific internal energy, $E_R$ the energy density of the radiation field, $\boldsymbol{H}$ the energy flux from non-ideal terms, $\boldsymbol{J}$ the current density, and $\boldsymbol{E}$ the electric field.
The non-ideal energy flux is:
\begin{equation}
\label{H}
\boldsymbol{H} = \boldsymbol{F}_R + \left(p_R + E_R \right) \boldsymbol{u} + \boldsymbol{Q} - \boldsymbol {\sigma}_{\nu} \cdot \boldsymbol {u}.
\end{equation}
where $\boldsymbol{F}_R$ is the radiative energy flux, $p_R$ is the radiation pressure and $\boldsymbol{Q}$ the heat flux.
Here, we have distinguished between the radiative enthalpy flux associated with the matter motion, $(E_R+p_R)\boldsymbol{u}$, and the radiative energy flux in the rest frame of the fluid, $F_R$ (see discussion in \cite{Shu1}).  

Differently from previous work, the above equations correctly describe quantum effects, which becomes important for high density fluids \citep{schmidt2012},  when the number density reaches values $\gtrsim 10^{24}$ cm$^{-3}$, as in white dwarfs or neutron star matter, or at small scales.
This means that Pauli blocking, tunnelling and wave packet spreading begin to exert an effective {\it quantum pressure} to the system \citep{quantum}.
This approach follows from the the fact that deterministic equations can be used to describe both single-particle and many-body distribution functions in the quantum limit if an appropriate potential is introduced in the hydrodynamic equations \citep{bohm, mostacci2008}.  See section 3.1 below for more detail.

\subsection{Induction Equation}
Starting from Ohm's law, and neglecting displacement current, we obtain:
\begin{multline}
\label{Beqn}
 {\partial \boldsymbol{B} \over \partial t} = \nabla \times(\boldsymbol{u} \times \boldsymbol{B}) + {\eta} \nabla^2 \boldsymbol{B} \\+ \frac{m}{e(1+Z)} \frac{\nabla p \times \nabla \rho}{\rho^2} + \nabla \times \left(\boldsymbol{B} \times{\tau_{ei} \over m_e}{\beta^{''}_1\chi^2+\beta^{''}_0 \over \Delta}\nabla T\right),
\end{multline}
where $\boldsymbol{B}$ is the magnetic field, $\eta$ the magnetic diffusivity ($\eta = 1 / \sigma_0 \mu_0$ where $\sigma_0$ is the electric conductivity and $\mu_0$ the vacuum permittivity), $m$ the average mass per particle, $m_e$ the electron mass, $e$ the elementary charge, $Z$ the degree of ionisation, $\tau_{ei}$ the electron-ion collision time, $\Delta = \chi^4 + \delta_1\chi^2 +\delta_0$ (where $\chi =\omega_{ce}\tau_{ei}$ is the Hall parameter and $\omega_{ce}$ is the electron cyclotron frequency), $T$ the fluid temperature, and $\beta^{''}_0$, $\beta^{''}_1$, $\delta_0$ and $\delta_1$ are Braginskii coefficients \citep{Braginskii1965}. Values for the Braginskii coefficients are given in Table 1.

\begin{table}[h]
	\centering
		\begin{tabular}{@{}lcccccl@{}} 
		\toprule[2pt]
		\centering
		\bf & \bf Z=1 &\bf Z=2 &\bf Z=3 &\bf Z=4 &\bf Z$\rightarrow \infty$   \\
		\midrule[1pt]
		$\beta^"_0$ 	&	3.053 &	1.784 &	1.442 &	1.285 & 	0.877	 \\
		$\delta_0$		&	3.7703 &	1.0465 &	0.5814 &	0.4106 &	0.0961 \\
		$\beta^"_1$     	&	1.5	&  	1.5	&	1.5	&	1.5	&	1.5	\\
		$\delta_1$		&	14.79	&	10.80	&	9.618	&	9.055	&	7.482	\\
		\bottomrule[2pt]
		\end{tabular}
		\begin{center}
	Table 1: {\bf Numerical values for the Braginskii coefficients for various values of Z, adapted from \citep{Braginskii1965}.} 
	 \end{center}
	
\end{table}

In addition to magnetic diffusion (second term on the right hand side of equation (\ref{Beqn})), we have written the induction equation to include baroclinic generation of magnetic field via the Biermann battery mechanism \citep{biermann,kulsrud}, and the advection of the magnetic field due to the Nernst effect \citep{HainesPPCF1986}. These are the last two terms on the right hand side of equation (\ref{Beqn}), respectively. In many laboratory and astrophysical scenarios, these terms represent the next highest order correction to Ohm's law \citep{Haines1986, Nishiguchi1985}.  While Ohm's law contains several additional terms \citep{HainesPPCF1986}, here we restrict to the case of small magnetic fields, where
zeroth order (Biermann battery) and first order (Nernst) terms have been shown to be the dominant mechanism for magnetic field generation in many plasma experiments \citep{Manuel2013}.  

\end{subequations}

\section{Optically thick source terms}
\label{ThickTerms}
Under the conditions of optically thick radiation, the source terms in equations (\ref{momentum}), (\ref{energy}) and (\ref{H}) are explicitly given by:

\begin{subequations}
\label{forms}
\begin{equation}
\label{radenflux}
-\nabla p_R = f_{rad}
\end{equation}
\begin{equation}
\label{radpress}
p_R ={ E_R \over 3} = { 4 \sigma T^4 \over 3 c} \\
\end{equation}
\begin{equation}
\label{Quantum}
\Phi_{Bohm} = -{\hbar^2 \rho \over 2 m_em_i} \nabla \left(\nabla^2 \sqrt{\rho} \over \sqrt{\rho} \right)
\end{equation}
\begin{equation}
\label{stress}
\boldsymbol{\sigma}_{\nu} = \rho \nu \left[ \nabla \boldsymbol{u} + (\nabla \boldsymbol{u} )^{T} - {2\over3}(\nabla \cdot \boldsymbol{u}) \boldsymbol{\underline{I}} \right] + \zeta(\nabla \cdot \boldsymbol{u})\underline{\boldsymbol{I}}
\end{equation}
\begin{equation}
\label{lorentz}
\boldsymbol{F}_{EM} = \rho_C \boldsymbol{E} + \boldsymbol{J} \times \boldsymbol{B} 
\end{equation}
\begin{equation}
\label{radflux}
\boldsymbol{F}_R = -{16 \sigma T^3 \over 3 \kappa_R \rho} \nabla T \\
\end{equation}
\begin{align}
\label{heattrans}
\boldsymbol{Q} &= - \kappa_{th} \nabla T = -\chi_{th} \rho c_p \nabla T \nonumber \\&= - \frac{\chi_{th} \rho k_B \gamma}{m (\gamma-1)} \nabla T
\end{align}
\end{subequations}
where $\sigma$ is the Stefan-Boltzmann constant, $c$ the speed of light, $\hbar$ the reduced Planck's constant, $m_i$ the ion mass, $\nu$ the kinematic viscosity with $\nu = \mu / \rho$, with $\mu$ being the (dynamic) viscosity, $\boldsymbol{\underline{I}}$ the identity tensor and $\zeta$ the second coefficient of viscosity, $\rho_C$ the charge density, $\kappa_R$ the Rosseland mean opacity, $\kappa_{th}$ the coefficient of heat conduction, $\chi_{th}$ the kinematic coefficient of thermal diffusivity, $c_p$ the specific heat capacity at constant pressure, $k_B$ Boltzmann's constant, and $\gamma$ the adiabatic index. 

When the fluid is optically thick, we can reduce the pressure tensor to a scalar radiation pressure and write it in terms of an isotropic energy density.  
Equation (\ref{radenflux}) is thus only applicable in this limit. Equation (\ref{radpress}) represents the isotropic thermal radiation pressure within the plasma, and the related energy density of that radiation assuming a Planck distribution \citep{radhydro}. 
Equation (\ref{radflux}) gives the radiative energy flux, within the local thermodynamic equilibrium (LTE) approximation.  In this form, it corresponds to the \emph{Rosseland heat flux} \citep{radhydro, drake2006high}. Clearly, these equations are not applicable to the case of optically thin systems, or when there is an optically thin pre-shock material, but optically thick post-shot material \citep{McClarren2010}.  We will discuss the optically thin case in section 6 below.

The quantum potential is explicitly given in Equation (\ref{Quantum}).  A derivation of this term is given in section \ref{Qpot}.  Equation (\ref{stress}) gives the form of the stress tensor. This forms does not assume that the fluid is incompressible, {\it i.e.} $\nabla \cdot \boldsymbol{u}$ does not have to be equal to zero \citep{drake2006high}.  It also considers the effects of viscosity to second order.  Equation (\ref{lorentz}) defines the electromagnetic (Lorentz) force on the system, in standard form. Finally, equation (\ref{heattrans}) describes the thermal heat flux in the diffusive limit \citep{landaufluid}.

\subsection{Quantum potential}
\label{Qpot}
Given the presence of the Bohm potential in the above equations, and the fact that this term is often omitted, it is important to give a detailed explanation and derivation of its appearance. The form used arises from rewriting the Schr\"{o}dinger equation in polar form with a wavefunction given by 
\begin{equation*} \phi = R e^{iS / \hbar},
\end{equation*} 
where $R$ and $S$ are real valued functions. The Schr\"{o}dinger equation can be thus divided into an imaginary part
\begin{equation}
\label{imaginary}
{\partial R \over \partial t} = -{1 \over 2m} \left(R\nabla^2S + 2\nabla R\cdot \nabla S \right),
\end{equation}
and a real part
\begin{equation}
\label{real}
{\partial S \over \partial t} = - \left[ {(\nabla S)^2 \over 2m} + V + Q \right],
\end{equation}
where $V$ is the external potential and 
\begin{equation*}
Q = -{\hbar^2 \over 2m}{\nabla^2 R \over R}.
\end{equation*}
If we now identify, using the correspondence to the  classical limit,  $R^2 = \rho$, and ${\bf u} = \nabla S/m$, then equation \ref{imaginary} can be re-expressed as a continuity
equation, while equation \ref{real} has the form of an energy equation with the classical potential corrected by the quantum term $Q$. This leads, for example, to the inclusion of 
$\rho Q /m$ as an energy density correction in the momentum equation.

In general the equations with quantum potential correction are written separately for the ion and electron species \citep{quantum}. 
For simplicity, we start by considering the case of an ideal fluid where the source terms are only pressure gradient and electromagnetic forces:

\begin{subequations}
\begin{multline}\label{mom_e} {\partial \boldsymbol{u}_e \over \partial t} + \boldsymbol{u}_e \cdot \nabla \boldsymbol{u}_e = -{\nabla p_e \over m_en_e} \\- {e \over m_e} (\boldsymbol{E} + \boldsymbol{u}_e \times \boldsymbol{B}) +  {\hbar^2 \over 2 m_e^2} \nabla \left( \nabla^2 \sqrt{n_e} \over \sqrt{n_e} \right),\end{multline}
\begin{multline} \label{mom_i}{\partial \boldsymbol{u}_i \over \partial t}+ \boldsymbol{u}_i \cdot \nabla \boldsymbol{u}_i = -{\nabla p_i \over m_in_i} \\+ {e \over m_i} (\boldsymbol{E} + \boldsymbol{u}_i \times \boldsymbol{B}) +  {\hbar^2 \over 2 m_i^2} \nabla \left( \nabla^2 \sqrt{n_i} \over \sqrt{n_i} \right),\end{multline}
\end{subequations}
where $n_e$ ($n_i$), $\boldsymbol{u_e}$ ($\boldsymbol{u_i}$), and $p_e$ ($p_i$) are the electron (ion) number density, velocity and pressure, respectively. Quantities with no subscript are instead used to describe average fluid properties.  By defining an average mass density and fluid velocity as \begin{equation*}\rho = m_en_e + m_in_i, \quad \boldsymbol{u} = {m_en_e\boldsymbol{u}_e + m_in_i\boldsymbol{u}_i \over  m_en_e + m_in_i},\end{equation*} 
we can combine equations \ref{mom_e}, 5b into a single fluid description 
by multiplying each one by $n_em_e$ and $n_im_i$, respectively, and by adding them together.  The resulting momentum equation is
\begin{multline*}
\rho \left({\partial \boldsymbol{u} \over \partial t}+ \boldsymbol{u} \cdot \nabla \boldsymbol{u}\right ) = -\nabla p + \boldsymbol{J} \times \boldsymbol{B} +  { \hbar^2 \rho\over 2 m_em_i} \nabla \left( \nabla^2 \sqrt{\rho} \over \sqrt{\rho} \right)\end{multline*}
which has the same form of the quantum potential as seen in equation (\ref{Quantum}).  In this derivation we have assumed quasi-neutrality, and taken the electron and ion pressures to be equal, that is $p_e = p_i = p / 2$.  This is correct except in the case of very large current densities.  

\section{Dimensionless Analysis}
We now rescale the variables in the hydrodynamic equations by a corresponding characteristic value.  This allows us to rewrite the equations in an invariant form, and the detail associated with the physical dimensions of the system is contained in a series of dimensionless numbers, which represent ratios of those characteristic values. 
We write the velocity, position, time and density as
\begin{eqnarray*}
&& \boldsymbol{u} \rightarrow u_0\boldsymbol{u}^{*},  \quad \boldsymbol{r} \rightarrow \ell_0\boldsymbol{r}^{*}, \quad t \rightarrow {\ell_0 \over u_0}t^{*}, \quad \rho \rightarrow \rho_0 \rho^{*},
\end{eqnarray*}
where $u_0$, $ \ell_0$, and $\rho_0$ are the characteristic velocity, length and density of the system, respectively. From now on we will use the convention that
\emph{starred quantities} ({\it i.e.} $\boldsymbol{u}^*$) are dimensionless, while \emph{quantities with subscript 0} ({\it i.e.} $u_0$) correspond to a characteristic value for that variable.  The above assumptions imply 
\begin{eqnarray*}
&& {\partial \over \partial t} \rightarrow {u_0 \over \ell_0} {\partial \over \partial t^{*}}, \quad \nabla \rightarrow {\nabla^{*} \over \ell_0}.
\end{eqnarray*}
Similarly, we can set
\begin{eqnarray*}
p \rightarrow p_0 p^{*}, \quad \boldsymbol{B} \rightarrow B_0\boldsymbol{B}^*, \quad \epsilon \rightarrow \epsilon_0 \epsilon^*.
\end{eqnarray*}
However, the choice of the values for $p_0$, $B_0$, and $\epsilon_0$ is not arbitrary.
To see this, we consider the momentum equation (\ref{momentum}), but with the only source terms being the pressure gradient and magnetic field ({\it i.e.}, in the ideal MHD limit). Using the relation:  $\boldsymbol{J} \times \boldsymbol{B} = (\boldsymbol{B} \cdot \nabla)\boldsymbol{B} / \mu_0 - \nabla (B^2/2 \mu_0)$ 
we can rewrite the momentum equation (\ref{momentum}) as
\begin{multline*}
\rho_0 \rho^* \left({u_0 \over \ell_0}{\partial u_0\boldsymbol{u}^* \over \partial t^*} + u_0\boldsymbol{u}^* \cdot  {\nabla \over \ell_0}^* u_0\boldsymbol{u}^*\right) \\=  -{\nabla \over \ell_0}^* p_0p^* + {B_0^{2}\over {\ell_0 \mu_0}} \left[(\boldsymbol{B}^* \cdot \nabla^*)\boldsymbol{B}^*  - \nabla^* \frac{B^{*2}}{2} \right].
\end{multline*}
Noticing the common factor of $u_0^2 \rho_0 / \ell_0$ on the left and dividing through gives
\begin{multline}
\rho^* \left({\partial \boldsymbol{u}^* \over \partial t^*} + \boldsymbol{u}^* \cdot  \nabla^* \boldsymbol{u}^*\right) =  -{p_0\over \rho_0u_0^2}\nabla^*p^* \\+ {B_0^{2} \over \mu_0 \rho_0 u_0^2} \left[(\boldsymbol{B}^* \cdot \nabla^*){\boldsymbol{B}^*} - \nabla^* {B^{*2} \over 2} \right].
\end{multline}
As we require this equation to have the same form as equation (\ref{momentum}), that is, to be invariant under the scaling transformation, this means that 
$p_0 \equiv \rho_0 u_0^2$ and $B_0 \equiv u_0 \sqrt{\mu_0 \rho_0}$. We see that the reference magnetic field has a value such that the fluid velocity and the Alfv\'{e}n velocity \citep{Alfven1942} are the same.

We can follow a similar procedure to determine the value for $\epsilon_0$.  Using the energy equation (\ref{energy}) in the ideal case with no source terms, we get
\begin{multline*}
{u_0 \over \ell_0}{\partial \over \partial t^*} \left( \rho_0 \epsilon_0 \rho^* \epsilon^* + {\rho_0 u_0^2 \rho^* u^{*2} \over 2} \right) \\= -  {\nabla \over \ell_0}^* \cdot \bigg[\rho_0 u_0  \rho^*\boldsymbol{u}^* \left(\epsilon_0 \epsilon^* + {u_0^2u^{*2} \over 2} \right) + \rho_0 u_0^2 p^* u_0\boldsymbol{u}^* \bigg].
\end{multline*}
Dividing through by a factor of $u_0^3 \rho_0 / \ell_0$ we obtain
\begin{multline*}
{\partial \over \partial t^*} \left( {\epsilon_0 \over u_0^2} \rho^* \epsilon^* + {\rho^* u^{*2} \over 2} \right) \\= - \nabla^* \cdot \left[\rho^*\boldsymbol{u}^* \left({\epsilon_0 \over u_0^2} \epsilon^* + {u^{*2} \over 2} \right) + p^* \boldsymbol{u}^* \right].
\end{multline*}

Again, we require this to be invariant under the scaling transformation, which leads to $\epsilon_0 \equiv u_0^2$.

This simple exercise has shown that the equations of ideal MHD are indeed invariant under scaling. This applies for any choice of the scaling transformation. In reality, the case is more complex because neither the laboratory system, nor the astrophysical one, can be assumed to always evolve under ideal conditions. To see this, we consider equation (\ref{Beqn}) with the inclusion of the resistive, baroclinic and Nernst terms. By applying the transformation defined above, with the additional inclusion of a temperature scaling
$T \rightarrow T_0 T^*$, we have
\begin{multline*}
 {u_0 \over \ell_0}\sqrt{\mu_0\rho_0}u_0{\partial \boldsymbol{B}^{*} \over \partial t^{*}} = - {u_0 \over \ell_0}\sqrt{\mu_0\rho_0}u_0 \nabla^{*} \times ( \boldsymbol{u}^{*} \times \boldsymbol{B}^{*}) \\+   {\eta} {u_0 \over \ell_0^2}.
\sqrt{\mu_0 \rho_0}u_0 \nabla^{*2} \boldsymbol{B}^{*} +
\frac{m u_0^2}{e \ell_0^2(1+Z)} \frac{\nabla^* p^* \times \nabla^* \rho^*}{\rho^{*2}}
\\+  {\sqrt{\mu_0 \rho_0}u_0T_0 \over \ell_0^2} {\tau_{ei} \over m_e}{\beta^{''}_1\chi^2+\beta^{''}_0 \over \Delta} \nabla^* \times \left(\boldsymbol{B}^* \times\nabla^* T^*\right).
\end{multline*}
Dividing through by \begin{math} u_0^2 \sqrt{\mu_0\rho_0} / \ell_0 \end{math}, 
\begin{multline}
 {\partial \boldsymbol{B}^{*} \over \partial t^{*}} = \nabla^{*} \times ( \boldsymbol{u}^{*} \times \boldsymbol{B}^{*})  + {1 \over Re_M} \nabla^{*2} \boldsymbol{B}^{*} \\+
\frac{1}{Bi} \frac{\nabla^* p^* \times \nabla^* \rho^*}{\rho^{*2}}
+{1 \over Ne} \nabla^* \times \left(\boldsymbol{B}^* \times\nabla^* T^*\right),
\end{multline}
where can recognize the magnetic Reynolds number as

\begin{equation*}
{1 \over Re_M} = {\eta \over u_0\ell_0},
\end{equation*}
which represents the ratio of magentic advection to magnetic diffusion,
and the dimensionless numbers
\begin{equation*}
{1 \over Bi} = \frac{m }{e \sqrt{\mu_0 \rho_0} \ell_0(1+Z)}, \quad {1 \over Ne} = {T_0 \tau_{ei} \over u_0 \ell_0 m_e}{\beta^{''}_1\chi^2+\beta^{''}_0 \over \Delta} 
\end{equation*}
which we will refer to as the Biermann number and Nernst number, respectively.  These numbers represent the importance of magnetic field generation, due to the presence of electron currents, relative to magnetic field advection.

This shows that the equations of resistive MHD are scale invariant only if $Re_M$, $Bi$ and $Ne$, are the same in both the laboratory and astrophysical systems, or, alternatively, very large in both systems, such that these terms are negligible.\\

\section{Similarity for non-ideal equations in the optically thick case}

We must now consider the full system of equations (\ref{energy})-(\ref{H}). In order to proceed, we
need to define additional scaling variables for current density, electric field and charge:
\begin{equation*}
\boldsymbol{J} \rightarrow J_0 \boldsymbol{J}^*, \quad \boldsymbol{E} \rightarrow E_0\boldsymbol{E}^*, \quad \rho_C \rightarrow \rho_{C_0} \rho_C^*.\\ \\
\end{equation*}

\subsection{Momentum equation}
We start with the momentum equation (\ref{momentum}) and use the above scaling transformations by dividing through a common factor 
$\rho_0u_0^2 / \ell_0$. After manipulation (for a more detailed derivation please see the Appendix) we get:

\begin{widetext}
\begin{multline}
\label{dimensionlessmomentum}
 \rho^* \left({\partial \boldsymbol{u}^* \over \partial t^*} + \boldsymbol{u}^* \cdot  \nabla^* \boldsymbol{u}^* \right) =
  -\nabla^*\left[ p^* + \frac{1}{R} T^{*4}\right]+ 
 \frac{1}{{\cal H}_Q} \rho^* \nabla^* \left({\nabla^{*2} \sqrt{\rho^*} \over \sqrt{\rho^*}} \right)
   \\+  \nabla^* \cdot \left\{ {1 \over Re} \bigg[ \nabla^* \boldsymbol{u}^* + (\nabla^* \boldsymbol{u}^* )^{T}  - {2\over3}(\nabla^* \cdot \boldsymbol{u}^*) \boldsymbol{\underline{I}} \bigg] + {1 \over Re_{\zeta}}(\nabla^* \cdot \boldsymbol{u}^*)\underline{\boldsymbol{I}} \right\}
   \\+ \frac{1}{\Omega_R} \rho_C^* \boldsymbol{E}^* + \frac{1}{\Omega_H} \boldsymbol{J}^* \times \boldsymbol{B}^* \\
\end{multline}
\end{widetext}
The Mihalas number ($R$) represents the ratio of ram pressure to radiation pressure, and it is related to the more familiar Boltzmann ($Bo$) number by
\begin{equation}
{1 \over R} = {{4 \sigma T_0^4 / 3 c}\over{\rho_0 u_0^2}} ={{4 u_0 \gamma}\over{3 c (\gamma-1) }}{1\over Bo},
\end{equation} 
where $Bo= \rho_0 c_p T_0 u_0 / \sigma T_0^4$.
Here, we have used $k_B T_0 \sim m u_0^2$, and $c_p \sim \gamma k_B / m (\gamma-1)$. The Boltzmann number
gives the ratio of the enthalpy flux with the radiation flux.

The importance of quantum effects against classical ones within the system is described by the number:
\begin{equation}  
{1 \over {\cal H}_Q} = {{\hbar^2} \over {2 m_e m_i \ell_0^2 u_0^2}},
\end{equation}
which we will refer to as the Bohm number.
We can also recognize the Reynold's number, the ratio of viscous to inertial effects, and its obvious extension when considering the second coefficient of viscosity:
\begin{equation}  {1 \over Re} = {\mu \over \rho_0 u_0 \ell_0}; \quad {1 \over Re_{\zeta}} = { \zeta \over \rho_0 u_0 \ell_0} \end{equation} \\
From charge conservation, $\rho_{C_0} = J_0 / u_0$, it follows
\begin{equation}
{1 \over\Omega_R} =  {\rho_{C_0} \ell_0 E_0 \over \rho_0 u_0^2}= {J_0 E_0 \ell_0 \over \rho_0 u_0^3},
\end{equation}
which represents the ratio between Ohmic and convective heat transfer. The ratio between convective transport and Hall diffusion is expressed by the coefficient
 \begin{equation}
{1 \over\Omega_H} =  {J_0 \ell_0 \over u_0 \sqrt{\mu_0\rho_0}}  = {J_0 B_0 \ell_0 \over {\mu_0\rho_0 u_0^2}}. \\
\end{equation}

\subsection{Energy equation}

Following the same approach as before, but now using the energy equation (\ref{energy}), and dividing through by a common factor of ${\rho_0 u_0^3 / \ell_0}$,  the dimensionless energy equation can thus be written as (see Appendix):
\begin{widetext}
\begin{multline}
\label{dimensionlessenergy}
 {\partial \over \partial t^*} \bigg(\rho^* \epsilon^* + {\rho^* u^{*2} \over 2} + \frac{3}{R} T^{*4} \bigg) + \nabla^* \cdot \bigg[\rho^* \boldsymbol{u}^* \left(\epsilon^* + {u^{2} \over 2} \right) + p^* \boldsymbol{u}^* \bigg] \\
 =  \nabla^* \cdot \left\{ -\frac{1}{\Pi_{thick}} {T^{*3} \over \rho^*} \nabla^* T^*  - \frac{3}{R} T^{*4} \cdot \boldsymbol{u}^* +  \frac{1}{Pe} \frac{\gamma}{\gamma - 1} \rho^* \nabla^* T^*  \right. 
 \\ \left. + \frac{1}{Re} \bigg[ \nabla^* \boldsymbol{u}^* + (\nabla^* \boldsymbol{u}^* )^{T} - {2\over3}(\nabla^* \cdot \boldsymbol{u}^*) \boldsymbol{\underline{I}} \bigg] \cdot \boldsymbol{u}^*+ \frac{1}{Re_\zeta} (\nabla^* \cdot \boldsymbol{u}^*)\underline{\boldsymbol{I}} \cdot \boldsymbol{u}^*  \right\} 
 \\-\frac{1}{\Omega_R} \boldsymbol{J}^*  \cdot \boldsymbol{E}^* + \frac{1}{{\cal H}_Q} \rho^* \nabla^* \left(\nabla^{*2} \sqrt{\rho^*} \over \sqrt{\rho^*} \right) \cdot \boldsymbol{u}^* - {1 \over R}T^{*4}\nabla^*\cdot \boldsymbol{u}^*
\end{multline}

\end{widetext}

Analogous to the momentum equation we have new dimensionless numbers. We define the radiation number, $\Pi_{thick}$, which is related to the Boltzmann number (in the same way as the Mihalas number, above) by:
\begin{equation}
\frac{1}{\Pi_{thick}} = {16 \sigma T_0^4 \over 3 \kappa_R \rho_0^{2} \ell_0 u_0^3} \left(= {16 \lambda_R \over 3 \ell_0}{\gamma \over \gamma -1}{1 \over Bo} \right).
\end{equation}
This number describes the importance of material energy flux compared to the radiative energy flux, weighted by the ratio of the mean free path of the radiation, $\lambda_R=1 / \kappa_R \rho_0$, to the characteristic length scale of the system.
The P\'{e}clet number gives the importance of thermal diffusion against convective transport:
\begin{equation}
\frac{1}{Pe} = {\chi_{th} k_B T_0 \over \ell_0 m u_0^3} = {\chi_{th} \over \ell_0  u_0 }, 
\end{equation}
where we have used again the relation $k_B T \sim m u_0^2$.

\section{Similarity for non-ideal equations in the optically thin case}
\label{ThinTerms}
It is worth noting that the Mihalas and the radiation numbers as shown above rely on the material in question being optically thick to radiation.  The form of the equations as formulated so far cannot be used in presence of optically thin radiation. The scaling relations in the optically thin case have been discussed in terms of cooling functions and characteristic timescales \citep{Livermore1999, Ryutov2001}, and using Lie group theory \citep{Falize2011a}. Moreover, in the special situation of thick-thin radiation transport a more complex treatment is required \citep{McClarren2010}.

Under optically thin conditions, the source terms relating to radiation can be written as
\begin{subequations}
\begin{equation}
\label{pRthin}
p_R = 0
\end{equation}
\begin{equation}
\label{fRadthin}
f_{rad} = 0
\end{equation}
\end{subequations}
where the transfer of momentum to the plasma by radiation is zero, by definition, as the plasma is optically thin, and the remaining radiation terms, relating to radiative energy flux (equation \ref{Cooling}), are written in terms of a cooling function,
\begin{equation}
\label{Cooling}
L_{\Lambda} = {\partial E_R \over \partial t} + \nabla \cdot \left[\boldsymbol{F}_R +E_R \boldsymbol{u}\right] \approx \rho \kappa_P \sigma T^4.
\end{equation}
This can be approximated with a form that is similar to the optically thick case (equation \ref{radpress}) where $\kappa_P$ is the Planck opacity.

The dimensionless momentum and energy equations now read as follows.
\begin{widetext}
\subsection{Momentum Equation}

\begin{multline}
\label{dimensionlessmomentumradthin}
 \rho^* \left({\partial \boldsymbol{u}^* \over \partial t^*} + \boldsymbol{u}^* \cdot  \nabla^* \boldsymbol{u}^* \right) =
  -\nabla^* p^* + 
 \frac{1}{{\cal H}_Q} \rho^* \nabla^* \left({\nabla^{*2} \sqrt{\rho^*} \over \sqrt{\rho^*}} \right)
   \\+  \nabla^* \cdot \left\{ {1 \over Re} \bigg[ \nabla^* \boldsymbol{u}^* + (\nabla^* \boldsymbol{u}^* )^{T}  - {2\over3}(\nabla^* \cdot \boldsymbol{u}^*) \boldsymbol{\underline{I}} \bigg] + {1 \over Re_{\zeta}}(\nabla^* \cdot \boldsymbol{u}^*)\underline{\boldsymbol{I}} \right\}
   \\+ \frac{1}{\Omega_R} \rho_C^* \boldsymbol{E}^* + \frac{1}{\Omega_H} \boldsymbol{J}^* \times \boldsymbol{B}^* \\
\end{multline}
\end{widetext}

\begin{widetext}
\subsection{Energy Equation}
\begin{multline}
\label{dimensionlessenergyradthin}
 {\partial \over \partial t^*} \bigg(\rho^* \epsilon^* + {\rho^* u^{*2} \over 2} \bigg) + \nabla^* \cdot \bigg[\rho^* \boldsymbol{u}^* \left(\epsilon^* + {u^{2} \over 2} \right) + p^* \boldsymbol{u}^* \bigg] \\
 =  \nabla^* \cdot \left\{ \frac{1}{Pe} \frac{\gamma}{\gamma - 1} \rho^* \nabla^* T^*  \right. 
 \left. + \frac{1}{Re} \bigg[ \nabla^* \boldsymbol{u}^* + (\nabla^* \boldsymbol{u}^* )^{T} - {2\over3}(\nabla^* \cdot \boldsymbol{u}^*) \boldsymbol{\underline{I}} \bigg] \cdot \boldsymbol{u}^*+ \frac{1}{Re_\zeta} (\nabla^* \cdot \boldsymbol{u}^*)\underline{\boldsymbol{I}} \cdot \boldsymbol{u}^*  \right\} 
 \\-\frac{1}{\Omega_R} \boldsymbol{J}^*  \cdot \boldsymbol{E}^* + \frac{1}{{\cal H}_Q} \rho^* \nabla^* \left(\nabla^{*2} \sqrt{\rho^*} \over \sqrt{\rho^*} \right) \cdot \boldsymbol{u}^* - {1\over \Pi_{thin}} \rho^* T^{*4}
\end{multline}

\end{widetext}

As for the optically thick case, we recover a set of dimensionless characteristic numbers. The only difference is that the radiation term is now altered, and the proper number to use in this case is
\begin{equation*}
{1 \over \Pi_{thin}}={\kappa_p \sigma \ell_0 T_0^4 \over u_0^3} = {\ell_0 \over \lambda_P}  {\sigma T_0^4 \over \rho_0 u_0^3} \left(= {\ell_0 \over \lambda_P} {\gamma \over \gamma -1} {1 \over Bo} \right)
\end{equation*}
which has a similar form to the radiation number for the optically thick case. It is a measure of the ratio between the material and radiative energy fluxes, weighted by the ratio of the mean free path, $\lambda_P = 1 / \kappa_P \rho_0$, to the characteristic length scale of the system $\ell_0$. However, please note that ratio between the radiation mean free path and $\ell_0$ is reversed when going from the optically thick to the optically thin regime.

\section{Discussion}

A summary of the scaling variables and all the dimensionless numbers is given in Table 2. As discussed earlier, similarity between the laboratory and astrophysical object is achieved if the dimensionless numbers are the same or suffiently large in both systems (the ideal MHD case). 
Under either of these conditions, we take  $\ell_0^{(1)}$, $u_0^{(1)}$, $\rho_0^{(1)}$, $J_0^{(1)}$, $E_0^{(1)}$, and
$T_0^{(1)}$ as the characteristic scaling parameters for the laboratory experiment. The astrophysical system has corresponding values given by
\begin{eqnarray}
\ell_0^{(2)} = g_a \ell_0^{(1)}, \quad
u_0^{(2)} = g_b u_0^{(1)}, \quad
\rho_0^{(2)} = g_c \rho_0^{(1)}, \nonumber \\ \nonumber
J_0^{(2)} = g_d J_0^{(1)}, \quad
E_0^{(2)} = g_e E_0^{(1)}, \quad
T_0^{(2)} = g_f T_0^{(1)},
\end{eqnarray}
where $g_{a,b,c,d,e,f}$ are scaling constants. From this set of parameters, we can scale all the other characteristic quantities as
\begin{eqnarray}
t_0^{(2)} = \frac{g_a}{g_b} t_0^{(1)}, \quad
p_0^{(2)} = g_c g_b^2 p_0^{(1)}, \quad
B_0^{(2)} = g_b\sqrt{g_c} B_0^{(1)}, \nonumber \\ \nonumber
\epsilon_0^{(2)} = g_b^2 \epsilon_0^{(1)}, \quad
\rho_{C_0}^{(2)} = \frac{g_d}{g_b} \rho_{C_0}^{(1)}.
\end{eqnarray}
All the details concerning the microphysics of the two systems are thus contained only in the dimensionless numbers given in Table 2. 
\begin{table}[h]
	\centering
		\begin{tabular}{@{}lcl@{}} 
		\toprule[2pt]
		\centering
		\bf Characteristic quantity & \bf Definition  \\
		\midrule[1pt]
		Length		 			&  $\ell_0$			 \\
		Velocity			 		&  $u_0$								 \\
		Density			 		&  $\rho_0$						 \\
		Current density 			&  $J_0$					\\		
		Electric field				&  $E_0$			 \\
		Temperature				&  $T_0$				    \\	
		\midrule[0.1pt]
		Time	 					&  $t_0 = \ell_0/u_0$			 \\
		Pressure		 			&  $p_0 = \rho_0 u_0^2$				 \\
		Magnetic field				& $B_0 = u_0 \sqrt{\mu_0\rho_0}$			 \\
		Specific internal energy		&  $\epsilon_0 = u_0^2$  \\		 
		Charge density 			&  $\rho_{C_0}=J_0/u_0$			 \\
		\midrule[0.1pt]
		Reynolds number			&  $Re = \rho_0 u_0 \ell_0 / \mu$\\
		Reynolds number (bulk)	 	&  $Re_\zeta = \rho_0 u_0 \ell_0 / \zeta$	 \\
		Magnetic Reynolds number 	&  $Re_M = u_0 \ell_0 / \eta$						\\
		Biermann number			&  $Bi = e (1+Z)\sqrt{\mu_0\rho_0} \ell_0 / m $ \\ 
		Nernst number				&  $Ne = u_0\ell_0 m_e \Delta / T_0 \tau_{ei} (\beta^{''}_1\chi^2+\beta^{''}_0) $ \\
		Mihalas number			&  $R = 3 c \rho_0 u_0^2/ 4 \sigma T_0^4$	 \\
		Radiation number	(Thick)		&  $\Pi_{thick} = 3 \ell_0 \rho_0  u_0^3/ 16 \lambda_R \sigma T_0^4$	 \\
		Radiation number	(Thin)		&  $\Pi_{thin} = \lambda_P \rho_0 u_0^3 / \ell_0 \sigma T_0^4$	 \\
		P\'eclet number				&  $Pe = \ell_0 u_0 / \chi_{th}$ \\
		Ohmic number				&  $\Omega_R = \rho_0 u_0^3 / J_0 E_0 \ell_0$ \\
		Hall number				&  $\Omega_H = \mu_0\rho_0 u_0^2 / J_0 B_0 \ell_0$ \\
		Bohm number			&  ${\cal H}_Q = 2 m_em_i u_0^2 \ell_0^2 / \hbar^2$ \\
		\bottomrule[2pt]
		\end{tabular}
		\begin{center}
	Table 2: {\bf List of scaling variables and dimensionless numbers} 
	 \end{center}
	
\end{table}
In order to evaluate these numbers, let's assume the plasma is in thermodynamic equilibrium at temperature $T$ (in eV) and carries a mass density 
$\rho$ (in g/cm$^3$) from ions of atomic mass $A$ and charge $Z$. The magnetic field is $B$ (in G). Charge neutrality implies an equal number of negative charges carried by mobile electrons. These assumptions are applicable to both the laboratory and astrophysical plasmas. Following \cite{Livermore1999,huba}, the kinematic viscosity is
\begin{equation}
	\nu \, ({\rm cm^2/s}) = {\rm Min}
	\left\{\begin{array}{ll}
	3.3\times 10^{-5} \, \frac{A^{1/2} T^{5/2}}{Z^4 \rho \Lambda}	\\
	2.8\times 10^{43} \, \frac{\rho^2 \Lambda}{A^{5/2} Z^2 B^2 T^{1/2}} 
	\end{array}\right\},
\end{equation} 
where $\Lambda$ is the Coulomb logarithm. The thermal diffusivity is \citep{Livermore1999} is
\begin{equation}
	\chi_{th} \, ({\rm cm^2/s}) = {\rm Min}
	\left\{\begin{array}{ll}
	3.3\times 10^{-3} \, \frac{A T^{5/2}}{Z (Z+1)\rho \Lambda}	\\
	8.6\times 10^{9} \, \frac{A^{1/2} T}{Z B}
	\end{array}\right\}.
\end{equation} 
The magnetic diffusivity can be written as \citep{landauPK}
\begin{equation}
	\eta \, ({\rm cm^2/s}) = 2.4\times 10^5 \, \frac{Z \Lambda}{T^{3/2}}.
\end{equation}
The electron-ion collision time is given by \citep{huba}
\begin{equation}
	\tau_{ei}  \, ({\rm s}) =  5.2\times 10^{-16} \, \frac{A^2 T^{3/2}}{Z^2 \rho \Lambda}.
\end{equation}
In the case of a fully ionized plasma, the Rosseland opacity is only determined by the free-free absorption, thus \citep{zeldovich2002} 
\begin{equation}
	\kappa_R \, ({\rm cm^2/g}) = 4.4 \times 10^8 \frac{Z^3 \rho}{A^2 T^{7/2}}.
\end{equation}
For typical astrophysical plasmas, the Planck opacity is \citep{sutherland}
\begin{equation}
	\kappa_P \, ({\rm cm^2/g}) = 1.8 \times 10^{13} \,\frac{Z \rho}{A^2 T^4},
\end{equation}
and for bremsstrahlung-dominated cooling \citep{Livermore1999}
\begin{equation}
	\kappa_P \, ({\rm cm^2/g}) = 3.1 \times 10^{10} \,\frac{Z^2 \rho}{A^2 T^{7/2}}
\end{equation}
At higher densities (near and above solid) and when line radiation transport must be included in the calculations, the Rosseland and Planck opacity are tabulated as \citep{Rosseland1987}
\begin{equation}
\label{rosselandeqn}
\kappa_{P, R} \, ({\rm cm^2/g}) = \kappa_0 \rho^\alpha T^\beta,
\end{equation}
where $\kappa_0$, $\alpha$ and $\beta$ are material dependent constants (see Tables 3 \& 4). The Rosseland and Planck opacities are bound to a maximum value given by \citep{Rosseland1987}
\begin{equation}
	\kappa_{P, R}^{\rm max} \, ({\rm cm^2/g}) = 6.1 \times 10^6 \, \frac{Z}{A T}.
\end{equation}

Even in the case that the dimensionless numbers are large in both the laboratory and astrophysical systems, their magnitude can be very different. It is then important to quantify the error in fluid variables in the ideal MHD approximation due to finite values for such dimensionless numbers. For the optically thick case, we have:
\begin{equation}
\frac{\Delta B}{B_{\rm id}} \sim \left(\frac{1}{Re_M^2}+\frac{1}{Bi^2}+\frac{1}{Ne^2}\right)^{1/2},
\end{equation}
\begin{eqnarray}
\frac{\Delta \rho u}{(\rho u)_{\rm id}} \sim \left(\frac{1}{R^2}+\frac{1}{{\cal H}^2_Q}+\frac{1}{Re^2}+\frac{1}{Re_\zeta^2}\right.+\\ \nonumber
\left.\frac{1}{\Omega_R^2}+\frac{1}{\Omega_H^2}\right)^{1/2},
\end{eqnarray}
\begin{eqnarray}
\frac{\Delta \rho \epsilon}{(\rho \epsilon)_{\rm id}} \sim \left(\frac{1}{R^2}+\frac{1}{\Pi_{thick}^2}+\frac{1}{Pe^2}\right.+\\ \nonumber
\left.\frac{1}{{\cal H}^2_Q}+\frac{1}{Re^2}+\frac{1}{Re_\zeta^2}+\frac{1}{\Omega_R^2}\right)^{1/2},
\end{eqnarray}
where $B_{\rm id}$, $(\rho u)_{\rm id}$, and $(\rho \epsilon)_{\rm id}$ refers to the magnetic field, momentum and energy, respectively,
in the ideal MHD approximation. Similar relations can be straightforwardly derived for optically thin plasmas.

\FloatBarrier

\begin{table}[h]
	\centering
		\begin{tabular}{@{}cccc@{}} 
		\toprule[2pt]
		\centering
		\bf Material & \bf $\kappa_0$ & \bf $\alpha$  & \bf $\beta$ \\
		\midrule[1pt]
		CH 	& $2.00\times 10^6$ 	& 0.14	& -2.00\\
		Al 	& $1.04\times 10^8$ 	& 0.48 	& -2.48 \\
		Ti 	& $3.07\times 10^7$ 	& 0.39 	& -2.21 \\
		Fe 	& $6.29\times 10^7$ 	& 0.31 	& -2.27 \\
		Cu 	& $5.93\times 10^7$ 	& 0.29 	& -2.21 \\
		Mo 	& $1.99\times 10^6$ 	& 0.22 	& -1.49 \\
		Sn 	& $3.70\times 10^6$ 	& 0.16 	& -1.57 \\
		Xe	& $2.00\times 10^8$ 	& 0.00 	& -2.00\\
		Ba 	& $5.89\times 10^6$ 	& 0.14 	& -1.62 \\
		Eu 	& $2.89\times 10^6$ 	& 0.09 	& -1.45 \\
		W 	& $5.59\times 10^5$ 	& 0.01 	& -1.12 \\	
		Au	& $6.00\times 10^6$ 	& 0.30	& -1.50\\		
		Pb 	& $4.11\times 10^5$ 	& 0.00 	& -1.05 \\
		U 	& $7.76\times 10^5$ 	& 0.04 	& -1.14 \\				
		\bottomrule[2pt]
		\end{tabular}
		\begin{center}
	Table 3: {\bf List of coefficient values for Rosseland opacity from equation (\ref{rosselandeqn}).  Adapted from \cite{Rosseland1987} and \cite{drake2006high}.}
	 \end{center}
\end{table}
\begin{table}[h]
	\centering
		\begin{tabular}{@{}cccc@{}} 
		\toprule[2pt]
		\centering
		\bf Material & \bf $\kappa_0$ & \bf $\alpha$  & \bf $\beta$ \\
		\midrule[1pt]
		CH 	& $2.00\times 10^5$ 	& 0.00	& -1.00\\
		Al 	& $6.01\times 10^8$ 	& 0.48 	& -2.42 \\
		Ti 	& $1.40\times 10^8$ 	& 0.44 	& -2.07 \\
		Fe 	& $2.22\times 10^8$ 	& 0.38 	& -2.13 \\
		Cu 	& $2.31\times 10^8$ 	& 0.36 	& -2.22 \\
		Mo 	& $1.54\times 10^7$ 	& 0.31 	& -1.56 \\
		Sn 	& $1.91\times 10^7$ 	& 0.23 	& -1.59 \\
		Xe	& $3.00\times 10^9$ 	& 0.00 	& -2.00\\
		Ba 	& $2.77\times 10^7$ 	& 0.24 	& -1.64 \\
		Eu 	& $1.68\times 10^7$ 	& 0.24 	& -1.54 \\
		W 	& $3.06\times 10^6$ 	& 0.20 	& -1.23 \\	
		Au	& $3.33\times 10^6$ 	& 0.17	& -1.23\\		
		Pb 	& $4.17\times 10^6$ 	& 0.16 	& -1.27 \\
		U 	& $1.04\times 10^7$ 	& 0.19 	& -1.42 \\	
		\bottomrule[2pt]
		\end{tabular}
		\begin{center}
	Table 4: {\bf List of coefficient values for Planck opacity from equation (\ref{rosselandeqn}).  Adapted from \cite{Rosseland1987} and \cite{drake2006high}.}
	 \end{center}
\end{table}

\section{Experimental Comparison}
In this section we apply the scaling relations to a few recent experiments and discuss how they can be used to meaningfully describe astrophysical environments. We focus our attention to the case when radiation becomes important, mainly because, as we will see below, this is where similarity between the laboratory and the astrophysical systems is difficult to achieve. On the other hand, in absence of significant radiative effects, hydrodynamic or MHD similarities has been successfully applied to wide range of problems. A comprehensive review of laboratory astrophysics experiments is given by \cite{Remington1999,Remington2006,drake2006high,savin2012}. 

Firstly, we consider an implosion experiment on the National Ignition Facility (NIF) laser \citep{Pak2013} used to model shock breakout in a circumstellar medium \citep{SN1993J}. While, as shown in Table 5, the experiments can indeed reproduce the supernova shock breakout in most aspects, the similarity breaks down when considering the Mihalas and radiation numbers. This means that the radiation pressure is significantly smaller than the material pressure and, in the laboratory, it does not change the form of the energy equation. Even if $R \gg 1$ in the laboratory, radiation can still be important in the energy equation, but also in this case, the large difference in the radiation number makes the similarity marginally satisfied (so the material energy flux is still larger than the radiative energy flux).
This example shows that radiation dominated environments are yet challenging to achieve even on the currently available largest laser facilities.  

Radiative jets and outflows are present in several young stellar objects \citep{reipurth2001}. Amongst more recent work, we focus on the
the experiment by \cite{Tikhonchuk2008}, who claim to have entered a regime where radiative effects are important. The scaling relations and corresponding dimensionless numbers are given in Table 6. We indeed see that in this specific case, the radiation number is significantly less than unity, though still many orders of magnitude different than in the case of stellar outflows.  Moreover, the Mihalas number still remains large in the laboratory, and, as before, full similarity breaks down. Small radiation numbers have been also achieved by \cite{Krauland2013a, Krauland2013b}, thus demonstrating that regimes where radiative flux, but not radiation pressure, is important, can be successfully scaled to astrophysical accretion shocks.  It is also important to note that, while the effects of radiative flux can be seen to dominate over material flux in the laboratory, the ratio is still many orders of magnitude different to the astrophysical case.  A well scaled experiment would have, at least, the correct direction of the ratio of the characteristic value ({\it i.e.}, large, if the value for the astrophysical case is large, or vice versa) for all quantities.

\begin{table}[h]
	\centering
		\begin{tabular}{@{}lcc@{}} 
		\toprule[2pt]
		\centering
		\bf Characteristic quantity & \bf Lab  & \bf Astro\\
		\midrule[1pt]
		Length		 			& 100 $\mu$m	&2.1 $\times 10^{10}$ km	 \\
		Velocity			 		&  	300 km/s	&		24,000 km/s			 \\
		Density			 		&  1 g/$\rm cm^3$			&	3.2 $\times 10^{-16}$  g/$\rm cm^3$		 \\
		Temperature				&  250 eV				&    86 keV\\	
		\midrule[0.1pt]
		Time	 					&  3 ps			& 10 days\\
		Pressure (Ram)		 			& 	90 TPa		& 190 Pa	 \\
		\midrule[0.1pt]
		Reynolds number			& 4.0 $\times 10^6$ &520\\
		Magnetic Reynolds number 	&   	8.7		&	1.3 $\times 10^{24}$		\\
		Biermann number                         &  130 & 2.3 $\times 10^{12}$ \\
		Nernst number			&0.1 & 7.8 $\times 10^{-3}$ \\
		Mihalas number			&  5,000	&7.4 $\times 10^{-19}$ \\
		Radiation number	(Thick)		&0.5  &$7.1 \times 10^{-19}$	 \\
		P\'eclet number				&1,200  &5.1\\
		Bohm number			& 6.4 $\times 10^{19}$ &2.1 $\times 10^{56}$\\
		\bottomrule[2pt]
		\end{tabular}
		\begin{center}
	Table 5: {\bf Example of scaling under radiative conditions from the laboratory \citep{Pak2013} to a  supernova breakout shock \citep{SN1993J}.
	}
	 \end{center}
\end{table}
\begin{table}[h]
	\centering
		\begin{tabular}{@{}lcc@{}} 
		\toprule[2pt]
		\centering
		\bf Characteristic quantity & \bf Lab  & \bf Astro\\
		\midrule[1pt]
		Length		 			& 150 $\mu$m	&7.5 $\times 10^{10}$ km	 \\
		Velocity			 		&  	500 km/s	&		100 km/s			 \\
		Density			 		&  1 $\times 10^{-4}$ g/$\rm cm^3$			&	1 $\times 10^{-22}$  g/$\rm cm^3$		 \\
		Temperature				&  100 eV				&    1 eV\\	
		\midrule[0.1pt]
		Time	 					&  300 ps			& 8,700 days\\
		Pressure (Ram)		 			& 	25 GPa		& 1 nPa	 \\
		\midrule[0.1pt]
		Reynolds number			& 9,200  & 2.6$ \times 10^7$\\
		Magnetic Reynolds number 	&   	1.7		&	3.7 $\times 10^{14}$		\\
		Biermann number                         &  0.6 & 2.0 $\times 10^{9}$ \\
		Nernst number			& 1.9 $\times 10^{-4}$ & 50 \\
		Mihalas number			&  55	&2.2 $\times 10^{-10}$ \\
		Radiation number	(Thick)		&4.1 $\times 10^{-7}$  &$1.2 \times 10^{-33}$	 \\
		P\'eclet number				&3.0  &7 $\times 10^5$\\
		Bohm number			& 6.0 $\times 10^{16}$ &6.0 $\times 10^{49}$\\
		\bottomrule[2pt]
		\end{tabular}
		\begin{center}
	Table 6: {\bf Comparison of laboratory experiment to an astrophysical case (Herbig-Haro object), with good scaling of radiative effects.  From \citep{Tikhonchuk2008}.}
	 \end{center}
\end{table}

Another aspect of the scaling relations that is worth discussing is the importance of the Bohm potential.
Whilst this term is of no significance in the tenuous interstellar plasma, it can become important when considering small scales, or compact objects, particularly
for densities exceeding $10^{23} - 10^{29}$ cm$^{-3}$ \citep{quantum}, which are found, for example, in white dwarfs and neutron stars.  

This is particularly relevant when considering, for example, Kolmogorov turbulence \citep{Kolmogorov1991}. In the inertial range
$\rho u_\ell^3/\ell = \dot{\epsilon} = \rm constant$, where $u_\ell$ is the characteristic velocity at scale $\ell$, and $\dot{\epsilon}$ is the total power
injected into turbulence. Hence the characteristic eddy turnover rate at scale $\ell$ is $u_\ell/\ell \sim (\dot\epsilon/\rho)^{1/3} \ell^{-2/3}$. 
Quantum effects are expected to become important when $\hbar^2/2 m_e \ell^2 \sim m u_\ell^2$, which defines the scale
\begin{equation}
\ell_q  \simeq \left(\frac{\hbar^2}{2 m_e m}\right)^{3/8} \left(\frac{\rho}{\dot{\epsilon}}\right)^{1/4}.
\end{equation}
We also notice that the rate of viscous dissipation on a scale $\ell$ is given by $\nu/\ell^2$. Equating this to the eddy turnover rate, we determine the scale at which viscous dissipation becomes dominant:
\begin{equation}
\ell_\nu \simeq  \nu^{3/4} \left(\frac{\rho}{\dot{\epsilon}}\right)^{1/4},
\end{equation}
and quantum effects will lie within the inertial range if $\ell_q > \ell_\nu$, or 
\begin{equation}
\nu \, ({\rm cm^2/s})   < \frac{1.9 \times 10^{-4}}{A^{1/2}}.
\end{equation}
Since the viscosity decreases as function of the density, it is then obvious to expect quantum effects to become more important at higher densities.
Similar considerations apply to the resistive scale. 
If the above conditions are satisfied, we would expect some change in the structure of turbulence below the scale $\ell_q$.  This can become important when considering the fluid core of white dwarf stars \citep{Bildsten2001} as shown in Table 7.

\begin{table}[h]
	\centering
		\begin{tabular}{@{}lc@{}} 
		\toprule[2pt]
		\centering
		\bf Characteristic quantity & \bf Astro  \\
		\midrule[1pt]
		Length		 			& $10^{3}$ km		 \\
		Velocity			 		&  	50 km/s				 \\
		Density			 		&  $10^7$ g/$\rm cm^3$					 \\
		Temperature				&  10 keV				\\	
		\midrule[0.1pt]
		$\ell_q / \ell_\nu$			&	50 \\
		\bottomrule[2 pt]
		\end{tabular}
		\begin{center}
	Table 7: {\bf Typical parameters for white dwarf stars, adapted from \cite{Lai2001, Zingale2009}. }
	 \end{center}
\end{table}

\section{Concluding Remarks}
In this paper we have provided a comprehensive description of the MHD scaling in presence of quantum, resistive and radiative effects. The dimensionless form of these equations reveals a set of characteristic numbers that can be used to quantify the departure from the ideal fluid behavior.
The scale invariance properties of the MHD equations have been successfully exploited to describe astrophysical phenomena in a variety of laboratory experiments  \cite{Remington1999,drake2006high,savin2012}, and here we have provided a unified theoretical framework that is common to all these experiments and can be applied to the planning and analysis of future ones.   

\acknowledgements
The research leading to these results has received funding from the European Research Council under the European Community's Seventh Framework Programme (FP7/2007-2013) / ERC grant agreement no. 256973. Partial support from AWE {\it plc~}is also acknowledged.
The authors would like to thank the anonymous referee for important insights provided into the manuscript.

\FloatBarrier

%

\appendix
The derivation of the dimensionless form of the momentum and energy equations (in the radiative thick regime) is outlined here in detail.

\section{Momentum Equation}
\label{AppendixMom}
Considering each term separately, we have:
\begin{subequations}
\begin{equation}
\rho \left({\partial \boldsymbol{u} \over \partial t} + \boldsymbol{u} \cdot  \nabla \boldsymbol{u} \right) \rightarrow {\rho_0 u_0^2 \over \ell_0} \rho^* \left({\partial \boldsymbol{u}^* \over \partial t^*} + \boldsymbol{u}^* \cdot  \nabla^* \boldsymbol{u}^* \right)
\end{equation}
\begin{equation}
- \nabla \left(p+{4 \sigma T^4 \over 3 c} \right) \rightarrow - {\rho_0 u_0^2 \over \ell_0} \nabla^* \left( p^* + {4 \sigma T_0^4 \over 3 \rho_0 u_0^2 c} T^{*4} \right)
\end{equation}
\begin{equation}
-{\hbar^2 \rho \over 2 m^2} \nabla \left(\nabla^2 \sqrt{\rho} \over \sqrt{\rho} \right)  \rightarrow -{\rho_0 u_0^2 \over \ell_0}\left({\hbar  \over  \ell_0u_0 \sqrt{2} m }\right)^2 \rho^* \nabla^* \left({\nabla^{*2} \sqrt{\rho^*} \over \sqrt{\rho^*}} \right) 
\end{equation}
\begin{multline}
\nabla \cdot \left\{ \rho \nu \left[ \nabla \boldsymbol{u} + (\nabla \boldsymbol{u} )^{T} - {2\over3}(\nabla \cdot \boldsymbol{u}) \boldsymbol{\underline{I}} \right] + \zeta(\nabla \cdot \boldsymbol{u})\underline{\boldsymbol{I}} \right\} \\  \rightarrow    {\rho_0 u_0^2 \over \ell_0}  \nabla^* \cdot \left\{ {\mu\over \rho_0 u_0\ell_0} \bigg[ \nabla^* \boldsymbol{u}^* + (\nabla^* \boldsymbol{u}^* )^{T} - {2\over3}(\nabla^* \cdot \boldsymbol{u}^*) \boldsymbol{\underline{I}} \bigg] \right. 
\left. + {\zeta \over \rho_0 u_0\ell_0}(\nabla^* \cdot \boldsymbol{u}^*)\underline{\boldsymbol{I}} \right\}
\end{multline}
\begin{equation}
\rho_C \boldsymbol{E} + \boldsymbol{J} \times \boldsymbol{B} \rightarrow {\rho_0 u_0^2 \over \ell_0} \left({\rho_{C_0} \ell_0 E_0 \over \rho_0 u_0^2} \rho_C^* \boldsymbol{E}^* + {J_0 \ell_0 \over u_0 \sqrt{\mu_0\rho_0}} \boldsymbol{J}^* \times \boldsymbol{B}^* \right). \\
\end{equation}
\end{subequations}
If we divide through by the common term $\rho_0 u_0^2 / \ell_0$, we obtain equation (\ref{dimensionlessmomentum}).

\section{Energy Equation}
\label{AppendixEn}
Considering again each term separately as we have done for the momentum equation:

\begin{subequations}
\begin{equation}
{\partial \over \partial t} \left( \rho \epsilon + {\rho u^{2} \over 2} + { 4 \sigma T^4 \over  c}\right)  \rightarrow  {\rho_0 u_0^3 \over \ell_0}{\partial \over \partial t^*} \left(\rho^* \epsilon^* + {\rho^* u^{*2} \over 2} + {4 \sigma T_0^4 \over \rho_0 u_0^2 c} T^{*4} \right),
\end{equation}
\begin{equation}
\nabla \cdot \left[\rho \boldsymbol{u} \left(\epsilon + {u^{2} \over 2} \right) + p \boldsymbol{u} \right]  \rightarrow {\rho_0 u_0^3 \over \ell_0}\nabla^* \cdot \left[\rho^* \boldsymbol{u}^* \left(\epsilon^* + {u^{2} \over 2} \right) + p^* \boldsymbol{u}^* \right],
\end{equation}
\begin{equation}
\nabla \cdot \left(-{16 \sigma T^3 \over 3 \kappa_R \rho} \nabla T\right)  \rightarrow {\rho_0 u_0^3 \over \ell_0}\nabla^* \cdot \left(-{16 \sigma T_0^4 \over 3 \kappa_R \rho_0^{2} \ell_0 u_0^3}{T^{*3} \over \rho^*} \nabla^* T^* \right),
\end{equation}
\begin{equation}
- \nabla \cdot \left({ 3 \sigma T^4 \over c} \right)\cdot \boldsymbol{u}  \rightarrow -{\rho_0 u_0^3 \over \ell_0} \nabla^* \cdot \left({4 \sigma T_0^4 \over \rho_0 u_0^2 c} T^{*4} \right) \cdot \boldsymbol{u}^*,
\end{equation}
\begin{equation}
- \nabla \cdot\left[{\chi_{th} \rho k_B \gamma \over m (\gamma - 1)} \nabla T \right]  \rightarrow {\rho_0 u_0^3 \over \ell_0} \nabla^*  \cdot\left[-{\chi_{th} k_B T_0 \gamma \over \ell_0 m u_0^3 (\gamma-1)}  \rho^*  \nabla^* T^* \right]
\end{equation}
\begin{multline}
\nabla \cdot \left\{ \rho \nu \left[ \nabla \boldsymbol{u} + (\nabla \boldsymbol{u} )^{T} - {2\over3}(\nabla \cdot \boldsymbol{u}) \boldsymbol{\underline{I}} \right] + 
\zeta(\nabla \cdot \boldsymbol{u})\underline{\boldsymbol{I}} \right\} \cdot \boldsymbol{u} 
\\ \rightarrow  
 {\rho_0 u_0^3 \over \ell_0}  \nabla^* \cdot \left\{ {\mu\over \rho_0 u_0\ell_0} \left[ \nabla^* \boldsymbol{u}^* + (\nabla^* \boldsymbol{u}^* )^{T} - {2\over3}(\nabla^* \cdot \boldsymbol{u}^*) \boldsymbol{\underline{I}} \right] \right.
 \\+ \left. {\zeta \over \rho_0 u_0\ell_0}(\nabla^* \cdot \boldsymbol{u}^*)\underline{\boldsymbol{I}} \right\} \cdot \boldsymbol{u}^*
\end{multline}
\begin{equation}
\boldsymbol{J} \cdot \boldsymbol{E} \rightarrow  {\rho_0 u_0^3 \over \ell_0} {J_0 E_0 \ell_0 \over \rho_0 u_0^3} \boldsymbol{J}^* \cdot \boldsymbol{E}^*
\end{equation}
\begin{equation}
 -{\hbar^2 \rho \over 2 m^2} \nabla \left(\nabla^2 \sqrt{\rho} \over \sqrt{\rho} \right) \cdot \boldsymbol{u} \rightarrow  -{\rho_0 u_0^3 \over \ell_0} \left({\hbar \over u_0\ell_0\sqrt{2} m}\right)^2 \rho^* \nabla^* \left(\nabla^{*2} \sqrt{\rho^*} \over \sqrt{\rho^*} \right) \cdot \boldsymbol{u}^*
\end{equation}
\begin{equation} 
-{4 \sigma T^4 \over 3c}\nabla \cdot \boldsymbol{u} \rightarrow -{\rho_0 u_0^3 \over \ell_0}\left({4 \sigma T_0^4 \over 3c u_0^2 \rho_0}\right)T^{*4}\nabla^*\cdot \boldsymbol{u}^*
\end{equation}
\end{subequations}
The factor ${\rho_0 u_0^3 / \ell_0}$ has been isolated from each term.  If we divide through by this, we then obtain equation (\ref{dimensionlessenergy}). 

\section{Full Equations}
Here we give a full summary of all the dimensionless equation of magneto-quantum-resistive hydrodynamics: 

\subsection{Continuity Equation}

\begin{equation*}
\label{ContinuityFull}
{\partial \rho^* \over \partial t^*} + \nabla^* \cdot \rho^* \boldsymbol{u}^* = 0 ,\\
\end{equation*}

\subsection{Induction Equation}

\begin{equation*}
 {\partial \boldsymbol{B}^{*} \over \partial t^{*}} = \nabla^{*} \times ( \boldsymbol{u}^{*} \times \boldsymbol{B}^{*})  + {1 \over Re_M} \nabla^{*2} \boldsymbol{B}^{*} +
\frac{1}{Bi} \frac{\nabla^* p^* \times \nabla^* \rho^*}{\rho^{*2}}
+{1 \over Ne} \nabla^* \times \left(\boldsymbol{B}^* \times\nabla^* T^*\right)
\end{equation*}

\subsection{Momentum Equation (Optically Thick)}

\begin{multline*}
\label{MomentumFullThick}
 \rho^* \left({\partial \boldsymbol{u}^* \over \partial t^*} + \boldsymbol{u}^* \cdot  \nabla^* \boldsymbol{u}^* \right) =
  -\nabla^*\left[ p^* + \frac{1}{R} T^{*4}\right]+ 
 \frac{1}{{\cal H}_Q} \rho^* \nabla^* \left({\nabla^{*2} \sqrt{\rho^*} \over \sqrt{\rho^*}} \right)
   \\+  \nabla^* \cdot \bigg\{ {1 \over Re} \bigg[ \nabla^* \boldsymbol{u}^* + (\nabla^* \boldsymbol{u}^* )^{T}  - {2\over3}(\nabla^* \cdot \boldsymbol{u}^*) \boldsymbol{\underline{I}} \bigg] + {1 \over Re_{\zeta}}(\nabla^* \cdot \boldsymbol{u}^*)\underline{\boldsymbol{I}} \bigg\}
   + \frac{1}{\Omega_R} \rho_C^* \boldsymbol{E}^* + \frac{1}{\Omega_H} \boldsymbol{J}^* \times \boldsymbol{B}^* \\
\end{multline*}

\subsection{Momentum Equation (Optically Thin)}

\begin{multline*}
 \rho^* \left({\partial \boldsymbol{u}^* \over \partial t^*} + \boldsymbol{u}^* \cdot  \nabla^* \boldsymbol{u}^* \right) =
  -\nabla^* p^* +  \frac{1}{{\cal H}_Q} \rho^* \nabla^* \left({\nabla^{*2} \sqrt{\rho^*} \over \sqrt{\rho^*}} \right)
   \\+  \nabla^* \cdot \left\{ {1 \over Re} \bigg[ \nabla^* \boldsymbol{u}^* + (\nabla^* \boldsymbol{u}^* )^{T}  - {2\over3}(\nabla^* \cdot \boldsymbol{u}^*) \boldsymbol{\underline{I}} \bigg] + {1 \over Re_{\zeta}}(\nabla^* \cdot \boldsymbol{u}^*)\underline{\boldsymbol{I}} \right\}
   + \frac{1}{\Omega_R} \rho_C^* \boldsymbol{E}^* + \frac{1}{\Omega_H} \boldsymbol{J}^* \times \boldsymbol{B}^* \\
\end{multline*}

\subsection{Energy Equation (Optically Thick)}

\begin{multline*}
 {\partial \over \partial t^*} \bigg(\rho^* \epsilon^* + {\rho^* u^{*2} \over 2} + \frac{3}{R} T^{*4} \bigg) + \nabla^* \cdot \bigg[\rho^* \boldsymbol{u}^* \left(\epsilon^* + {u^{2} \over 2} \right) + p^* \boldsymbol{u}^* \bigg] \\
 =  \nabla^* \cdot \left\{ -\frac{1}{\Pi_{thick}} {T^{*3} \over \rho^*} \nabla^* T^*  - \frac{3}{R} T^{*4} \cdot \boldsymbol{u}^* +  \frac{1}{Pe} \frac{\gamma}{\gamma - 1} \rho^* \nabla^* T^*  \right. 
 \\ \left. + \frac{1}{Re} \bigg[ \nabla^* \boldsymbol{u}^* + (\nabla^* \boldsymbol{u}^* )^{T} - {2\over3}(\nabla^* \cdot \boldsymbol{u}^*) \boldsymbol{\underline{I}} \bigg] \cdot \boldsymbol{u}^*+ \frac{1}{Re_\zeta} (\nabla^* \cdot \boldsymbol{u}^*)\underline{\boldsymbol{I}} \cdot \boldsymbol{u}^*  \right\} 
 \\-\frac{1}{\Omega_R} \boldsymbol{J}^*  \cdot \boldsymbol{E}^* + \frac{1}{{\cal H}_Q} \rho^* \nabla^* \left(\nabla^{*2} \sqrt{\rho^*} \over \sqrt{\rho^*} \right) \cdot \boldsymbol{u}^* - {1 \over R}T^{*4}\nabla^*\cdot \boldsymbol{u}^*
\end{multline*}

\subsection{Energy Equation (Optically Thin)}

\begin{multline*}
 {\partial \over \partial t^*} \bigg(\rho^* \epsilon^* + {\rho^* u^{*2} \over 2} \bigg) + \nabla^* \cdot \bigg[\rho^* \boldsymbol{u}^* \left(\epsilon^* + {u^{2} \over 2} \right) + p^* \boldsymbol{u}^* \bigg] \\
 =  \nabla^* \cdot \left\{ \frac{1}{Pe} \frac{\gamma}{\gamma - 1} \rho^* \nabla^* T^*  \right. 
 \left. + \frac{1}{Re} \bigg[ \nabla^* \boldsymbol{u}^* + (\nabla^* \boldsymbol{u}^* )^{T} - {2\over3}(\nabla^* \cdot \boldsymbol{u}^*) \boldsymbol{\underline{I}} \bigg] \cdot \boldsymbol{u}^*+ \frac{1}{Re_\zeta} (\nabla^* \cdot \boldsymbol{u}^*)\underline{\boldsymbol{I}} \cdot \boldsymbol{u}^*  \right\} 
 \\-\frac{1}{\Omega_R} \boldsymbol{J}^*  \cdot \boldsymbol{E}^* + \frac{1}{{\cal H}_Q} \rho^* \nabla^* \left(\nabla^{*2} \sqrt{\rho^*} \over \sqrt{\rho^*} \right) \cdot \boldsymbol{u}^* - {1\over \Pi_{thin}} \rho^* T^{*4}
\end{multline*}

\end{document}